\documentclass[11pt]{article}
\usepackage{amsmath, amssymb}
\usepackage{graphics,color,booktabs}
\usepackage{easybmat}
\usepackage{multirow}
\renewcommand{\baselinestretch}{1.5}

\numberwithin{equation}{section}

\addtocounter{page}{0}
\newtheorem{thm}{Theorem}[section]

\newtheorem{lem}{Lemma}[section]
\allowdisplaybreaks

\begin{document}

\title{\bf Quantile Regression for Partially Linear Varying Coefficient Spatial Autoregressive Models}

\author{Xiaowen Dai$^1$, Shaoyang Li$^1$, Maozai Tian$^{1,2}$\\
\small 1. Center for Applied Statistics, School of Statistics, Renmin University of China, Beijing 100872, China \\
\small 2. School of Statistics, Lanzhou University of Finance and Economics Lanzhou, 730101, Gansu, China}
\maketitle

\begin{abstract}
This paper considers the quantile regression approach for partially linear spatial autoregressive models with possibly varying coefficients. B-spline is employed for the approximation of varying coefficients. The instrumental variable quantile regression approach is employed for parameter estimation. The rank score tests are developed for hypotheses on the coefficients, including the hypotheses on the non-varying coefficients and the constancy of the varying coefficients. The asymptotic properties of the proposed estimators and test statistics are both established. Monte Carlo simulations are conducted to study the finite sample performance of the proposed method. Analysis of a real data example is presented for illustration.  
\end{abstract}

\section{Introduction}

Spatial econometric models have been widely used in many areas (e.g., economics, political science and public health) to deal with spatial interaction effects among geographical units (e.g., jurisdictions, regions, and states). Many of the early studies have been summarized in Anselin (1988), Anselin and Bera
(1998), LeSage (1999) and LeSage and Pace (2009). Recently, there are a large number of literature concerning on the spatial econometric models. For instance, Lee (2007) studied the generalized method of moments (GMM)
applied to the Spatial autoregressive model. Lee (2004) studied asymptotic properties
of the quasi-maximum likelihood estimator of the Spatial autoregressive model. Lee and Yu (2010) proposed the maximum likelihood (ML) estimator for the spatial autoregressive (SAR) panel model with both spatial lag and spatial disturbances. Dai, et al. (2015, 2016) respectively studied the local influence and outlier detection in the general spatial model which includes the spatial autoregressive model and the spatial error model as two special cases. Xu and Lee (2015) considered the instrumental variable (IV) and MLE estimators for spatial autoregressive model with a nonlinear transformation of the
dependent variable. Qu and Lee (2015) provided three estimation methods for the spatial autoregressive model with an endogenous spatial weight matrix, including two-stage
instrumental variable (2SIV) method, quasi-maximum likelihood estimation (QMLE) approach, and
generalized method of moments (GMM). Zhang and Shen (2015) investigated the GMM estimation approach for the partially linear varying coefficient spatial autoregressive panel data models with random effects. Jin, et al. (2016) studied oulier detection in the spatial autoregressive model.

However, in some practical applications, a linear model might not be flexible enough to capture the underlying complex dependence structure. And a purely nonparametric model may suffer from the so-called ``curse of dimensionality'' problem, the practical implementation might not be easy, and the visual display may not be useful for the exploratory purposes. To deal with the aforementioned problems, some dimension reduction modeling methods have been proposed in the literature. For example, He et al. (1998), He and Ng (1999), He and Portnoy (2000), De Gooijer and Zerom (2003), Yu and Lu (2004) considered the additive quantile regression models for iid data. Honda (2004) and Cai and Xu (2008) proposed the varying coefficient quantile regression models for time series data. He and Shi (1996), He and Liang (2000), and Lee (2003) considered the partially linear quantile regression models for iid data. Ahmad, Leelahanon and Li (2005) and Fan and Huang (2005) considered the partially linear varying coefficient models for cross-sectional data. Sun and Wu (2005) and Fan, Huang and Li (2007) considered the partially linear varying coefficient models for longitudinal data.

In this paper, we investigate the quantile regression approach for partially linear varying coefficient spatial autoregressive models, since the partially linear varying coefficient model is a good balance between flexibility and parsimony. We employ B-spline for the approximation of varying coefficients. Due to the presence of endogenous variable, we employ the instrumental variable quantile regression (IVQR) method to attenuate the bias. The focus of this paper is to estimate the conditional quantile curves without any specification of the error distribution.

The rest of the paper is organized as follows. Section \ref{sec:pvsar} introduces the partially linear varying coefficient spatial autoregressive models. Section \ref{sec:method} proposes the IVQR estimation procedure. Section \ref{sec:test} proposes the inference procedures for testing the non-varying coefficients and the constancy of the varying coefficients. The asymptotic properties of the estimators and test statistics are also discussed. Proofs of the theorems in Sections \ref{sec:method} and \ref{sec:test} are given in the Appendix. Section \ref{sec:simu} reports a simulation study for assessing the finite sample performance of the proposed estimators. An empirical illustration is considered in Section \ref{sec:real}. Section \ref{sec:con} concludes the paper.

\section{The Models} \label{sec:pvsar}

Consider the following partially linear varying coefficient spatial autoregressive model
\begin{equation} \label{mod:01}
y_{i}=\rho\sum_{j=1}^{n}w_{ij}y_{j}+X^{\top}_{i}\beta+Z^{\top}_{i}\gamma(U_{i})+\varepsilon_{i},
\end{equation}
where $y_{i}$ is the dependent variable, $X_{i}$ is a $p\times1$ vector, $Z_{i}$ is
a $q\times1$ vector. $w_{ij}$ is the $(i,j)$th element of the spatial weight matrix $W$.
The parameter $\rho$ is a coefficient on the spatial lagged dependent variable $Wy$,
$\beta$ is a $p\times1$ parameter vector, $\gamma(U_{i})$ comprises $q$ unknown smooth functions,
$U_{i}$ is the smoothing variable. Here, we only consider one-dimensional smoothing variable $U_{i}\in\mathbb{R}$.

Matrix form of model (\ref{mod:01}) is
\begin{equation} \label{mod:02}
y=\rho Wy+X\beta+Z\gamma(U)+\varepsilon,
\end{equation}
where $y=(y_{1},\cdots,y_{n})^\top$, $X=[X_{1},\cdots,X_{n}]^{\top}$, $X_i=[X_{i1},\cdots,X_{ip}]^{\top}$, $Z=[e_1\otimes Z_{1}^\top,\cdots,e_n\otimes Z_{n}^\top]$, $Z_i=[Z_{i1},\cdots,Z_{iq}]^{\top}$, $U=(U_{1},\cdots,U_{n})^\top$, $e_i$ is an $n\times1$ vector with the $i$th element equal to 1 and the rest equal to 0, $\gamma(U)=(\gamma^{\top}(U_{1}),\cdots,\gamma^{\top}(U_{n}))^\top$ is an $nq\times1$ vector, $\varepsilon=(\varepsilon_{1},\cdots,\varepsilon_{n})^\top$. Here, we can denote $Z^*=[Z_1,\cdots,Z_n]^\top$.

Due to the presence of endogenous variable $d_{i}=\sum_{j=1}^{n}w_{ij}y_{j}$, we employ the instrumental variable quantile regression (IVQR) method to attenuate the bias. The endogenous variable $d_{i}$ is related to a vector of instruments $\omega_{i}$ which are independent of $\varepsilon_{i}$. Then we can define the following conditional instrumental quantile relationship:
\begin{equation} \label{mod:04}
Q_{\tau}(y_{i}|\mathcal{F}_{-i},X_{i},Z_{i},U_i)
=\rho(\tau)d_{i}+X^{\top}_{i}\beta(\tau)+Z^{\top}_{i}\gamma(\tau,U_{i})+\omega_{i}\zeta(\tau),
\end{equation}
where $Q_\tau({y_{i}}|\mathcal{F}_{-i},X_{i},Z_{i},U_i)$ is the conditional $\tau$-quantile of $y_{i}$ given $\mathcal{F}_{-i},X_{i},Z_{i}$ and $U_i$, $\mathcal{F}_{-i}$ is the $\sigma$-field of $\{y_j:j\neq i\}$, $\zeta$ is the coefficient corresponding to the instrumental variable $\omega_i$, $Q_{\tau}(\varepsilon_{i}|\mathcal{F}_{-i},X_{i},Z_{i},U_i)=0$.

\section{The proposed method} \label{sec:method}

\subsection{Instrumental Variable Quantile Regression Estimator (IVQR)}

In this section, we employ B-spline for estimation. Without loss of generality, we assume that $U_{i}\in[0,1]$ for all $i$ throughout.

We employ normalized B-splines of order $h+1$ to approximate the $\gamma_{l}(\tau,u)$, $l=1,\cdots,q$. We consider a sequence of positive integers $\{k_{n}\}$, $n\geq1$, and an extended partition of $[0,1]$ by $k_{n}$ quasi-uniform internal knots. Let $\pi_{k_{n}}(u)=(B_{1}(u),\cdots,B_{k_{n}+h+1}(u))^{\top}$ denote a set of B-spline basis functions. We approximate each $\gamma_{l}(\tau,u)$ by a linear combination of normalized B-spline basis functions
$$
\gamma_{l}(\tau,u)\approx\sum_{s=1}^{k_{n}+h+1}B_s(u)\theta_{l,s}(\tau)=\pi_{k_{n}}(u)^{\top}\theta_{l}(\tau),
$$
where $\theta_{l}(\tau)=(\theta_{l,1}(\tau),\cdots,\theta_{l,k_{n}+h+1}(\tau))^{\top}$ is the spline coefficient vector. For details
on the construction of B-spline basis functions, the readers are referred to Schumaker (1981). With the B-spline basis, model \eqref{mod:04} can be approximated by
\begin{align} \label{mod:05}
Q_{\tau}(y_{i}|\mathcal{F}_{-i},X_{i},Z_{i},U_i)&\approx\rho(\tau)d_{i}+\sum_{l=1}^{p}
X_{i,l}\beta_{l}(\tau)+\sum_{l=1}^{q}\sum_{s=1}^{k_{n}+h+1}Z_{i,l}B_{s}(U_{i})\theta_{l,s}(\tau)+\omega_{i}\zeta(\tau), \nonumber \\
&=\rho(\tau)d_{i}+X^{\top}_{i}\beta(\tau)+\Pi^{\top}_{i}\Theta(\tau)+\omega_{i}\zeta(\tau),
\end{align}
where $\Pi_{i}=(Z_{i1}\pi_{k_{n}}^{\top}(U_i),\cdots,Z_{iq}\pi_{k_{n}}^{\top}(U_i))^{\top}\in\mathbb{R}^{q_{k_{n}}}$, $\Theta(\tau)=(\theta_{l,s}(\tau))\in\mathbb{R}^{q_{k_n}}$, $q_{k_{n}}=q(k_{n}+h+1)$.

Then we can define the following objective function:
\begin{equation} \label{ob:iv}
R_{IV}(\tau,\rho,\beta,\Theta,\zeta)=\sum_{i=1}^{n}\rho_{\tau}(y_{i}-\rho d_i-X^{\top}_{i}\beta
-\Pi^{\top}_{i}\Theta-\omega_{i}\zeta).
\end{equation}
Following Chernozhukov and Hansen (2006, 2008) and Galvao (2011), and assuming the availability of instrumental variables $\omega_{i}$, we can derive the IVQR estimator via the following three steps:

\begin{itemize}
\item Step 1: For a given quantile $\tau$, define a suitable set of values $\{\rho_j,j=1,\cdots,J;|\rho|<1\}$. One then minimizes the objective function for $\beta,\Theta,\zeta$ to obtain the ordinary QR estimators
of $\beta,\Theta,\zeta$:
\begin{equation}
(\hat\beta(\rho,\tau),\hat\Theta(\rho,\tau),\hat\zeta(\rho,\tau))
=\underset{\beta,\Theta,\zeta}{\arg\min}R_{IV}(\tau,\rho,\beta,\Theta,\zeta).
\end{equation}

\item Step 2: Choose $\hat\rho(\tau)$ among $\{\rho_j,j=1,\cdots,J\}$ which makes a weighted distance function defined on $\zeta$ closest to zero:
\begin{equation}
\hat{\rho}(\tau)=\underset{\rho\in\mathcal{R}}{\arg\min}\bigg\{\hat\zeta(\rho,\tau)^{\top}
\hat{A}(\tau)\hat\zeta(\rho,\tau)\bigg\},
\end{equation}
where $A$ is a positive definite matrix, $\mathcal{R}=[-1,1]$.

\item Step 3: The estimation of $\beta,\Theta$ can be obtained, which is respectively $\hat{\beta}(\hat{\rho}(\tau),\tau)$ and $\hat{\Theta}(\hat{\rho}(\tau),\tau)$. Accordingly, the polynomial spline estimator $\hat\gamma_{l}(\tau,u)$ is given by $\pi_{k_{n}}(u)^{\top}\hat\theta_{l}(\tau)$ for each $l$, $l=1,\cdots,q$.
\end{itemize}

\textit{Remark 1}. Throughout the paper, we use the cubic spline in the B-spline approximation. For the objective function \eqref{ob:iv}, the knots $k_n$ are chosen as the minimizer to the following Schwarz-type Information Criterion:
\begin{align*}
SIC(k_n)&=\log\bigg\{\sum_{i=1}^{n}\rho_\tau\bigg(y_{i}-\hat\rho_{(k_n)}\sum_{j=1}^{n}w_{ij}y_{j}-X^{\top}_{i}
\hat\beta_{(k_n)}-\Pi^{\top}_{i}\hat\Theta_{(k_n)}-\omega_{i}\hat\zeta_{(k_n)}\bigg)\bigg\} \\
&+\frac{\log n}{2n}(2+p+q_{k_n}).
\end{align*}
where $\hat\rho_{(k_n)},\hat\beta_{(k_n)},\hat\Theta_{(k_n)},\hat\zeta_{(k_n)}$ are the $\tau$th quantile estimators with $k_n$ knots. More details can be found in Kim (2003).

\textit{Remark 2}. For an IVQR estimation, we need instruments for the endogenous variable $D=Wy$. In practice, we can choose $WX$, $[X,Z^*]$, $[WX,WZ^*]$, etc. as instrumental variable matrix. In this paper, $[WX,WZ^*]$ is chosen as instrumental variable matrix.

\subsection{Asymptotic theory}

The following are sufficient conditions for the proposed IVQR estimator based on polynomial spine approximation.

\textbf{Assumption 1}

(i) $(y_i,U_i,D_i)$ are independent and identically distributed (i.i.d.) for each fixed $i$ with conditional distribution function $F$ for $i=1,\cdots,n$.

(ii) The conditional distribution of $U$ given $Z=z$ has a bounded density $f_{U|Z}$, which satisfies $0<c_{1}\leq f_{U|Z}(u|z)\leq c_{2}<\infty$ uniformly in $z$ and $u$ for some constants $c_{1},c_{2}>0$.


(iii) Uniformly over $i$, $\varepsilon_{i}$ has a bounded density function $f_{i}$ that is continuously differentiable in the neighbourhood of 0 with first derivative bounded.

\textbf{Assumption 2}

(i) $\gamma_l(u)\in\mathcal{V}$, where $\mathcal{V}$ denotes the class of varying coefficient functions. For some $r\geq1$, $\gamma_{l}(u)\in\mathcal{H}_{r}$, $l=1,\cdots,q$.

Here, we say function $g(u)$ belongs to the class of varying coefficient functions $\mathcal{V}$ if $g(u)=z^\top h(u)$ and $E\|g(u)\|^2<\infty$. And $\mathcal{H}_{r}$ denote the collection
of all functions on $[0,1]$ whose $m$th order derivative satisfies the H\"{o}lder condition of order $\nu$ with $r\equiv m+\nu$. That is, for any $h\in\mathcal{H}_{r}$, $|h^{(m)}(s)-h^{(m)}(t)|\leq c|s-t|^{\nu}$, for any $s,t\in[0,1]$ and $c>0$.

(ii) For any varying coefficient function $g(u)$ defined on $\mathcal{U}$, $\sup_{u\in\mathcal{U}}\|g(u)-z^\top h(u)\|=O((k_n+h+1)^{-r})$.

\textbf{Assumption 3}

(i) For all $\tau\in\mathcal{T}$, $(\rho(\tau),\beta(\tau),\Theta(\tau,u))$ is in the interior of the set $\mathcal{R}\times\mathcal{B}\times\mathcal{S}$, and $\mathcal{R}\times\mathcal{B}\times\mathcal{S}$ is compact and convex.

(ii) Let
\begin{align}
\Phi(\rho,\beta,\Theta,\zeta,\tau)&=\mathbb{E}[(\tau-I(y<D\rho+X\beta+\Pi\Theta+E\zeta))
\tilde{X}],\\
\Phi(\rho,\beta,\Theta,\tau)&=\mathbb{E}[(\tau-I(y<D\rho+X\beta+\Pi\Theta))\tilde{X}],
\end{align}
where $\tilde{X}=[X,\Pi,E]$, $D=Wy$, $E=(\omega_1,\cdots,\omega_n)^\top$. The Jacobian matrices $\frac{\partial\Phi(\rho,\beta,\Theta,\tau)}{\partial(\rho,\beta,\Theta)}$ and $\frac{\partial\Phi(\rho,\beta,\Theta,\zeta,\tau)}{\partial(\beta,\Theta,\zeta)}$ are continuous and have full rank uniformly over $\mathcal{R}\times\mathcal{B}\times\mathcal{S}\times\mathcal{Z}\times\mathcal{T}$. The parameter space $\mathcal{R}\times\mathcal{B}\times\mathcal{S}$ is a connected set and the image of $\mathcal{R}\times\mathcal{B}\times\mathcal{S}$ under the map $(\rho,\beta,\Theta)\mapsto\Phi(\rho,\beta,\Theta,\tau)$ is simply connected.

(iii) Denote $\Omega=\text{diag}(f_{i}(\xi_{i}(\tau)))$, where $\xi_{i}(\tau)=\rho(\tau)d_{i}+X^{\top}_{i}\beta(\tau)+\Pi^{\top}_{i}\Theta(\tau)
+\omega_{i}\zeta(\tau)$. Let $\eta=(\beta^\top,\Theta^\top,\zeta)^\top$. Then, the following matrices are positive definite:
\begin{align}
\mathbf{J}_{\eta}&=\underset{n\rightarrow\infty}{\lim}\frac{1}{n}\tilde{X}^{\top}\Omega\tilde{X}, \\ \mathbf{J}_{\rho}&=\underset{n\rightarrow\infty}{\lim}\frac{1}{n}\tilde{X}^{\top}\Omega D,  \\
S&=\underset{n\rightarrow\infty}{\lim}\frac{\tau(1-\tau)}{n}\tilde{X}^\top\tilde{X}.
\end{align}
Let $[\bar{\mathbf{J}}^\top_{\beta},\bar{\mathbf{J}}^\top_{\Theta},\bar{\mathbf{J}}_\zeta^\top]$ be a
conformable partition of $\mathbf{J}^{-1}_{\eta}$ and $H=\bar{\mathbf{J}}^\top_\zeta A\bar{\mathbf{J}}_\zeta$. Hence, $\mathbf{J}_{\eta}$ is invertible and $\mathbf{J}_{\rho}^{\top}H\mathbf{J}_{\rho}$ is also invertible.

(iv) $\max\|y_{i}\|=O(\sqrt{n})$, $\max\|X_{i}\|=O(\sqrt{n})$, $\max\|Z_{i}\|=O(\sqrt{n})$, $\max\|\omega_{i}\|=O(\sqrt{n})$, and $\max\|\Pi_i\|=O(\sqrt{k_n+h+1})$.

\begin{thm}[Uniformly Convergence] \label{th:1} Under Assumptions 1-3, $\rho(\tau),\beta(\tau),\Theta(\tau)$ are consistently estimable. And if $r\geq1$, then
$$
\sup_{l\in\{1,\cdots,q\}}\sup_{u\in\mathcal{U}}\|\hat\gamma_l(u,\tau))-\gamma_l(u,\tau)\|=O_p((k_n+h+1)^{-r}).
$$
\end{thm}

\begin{thm}[Asymptotic Distribution] \label{th:2}
(i) Under Assumptions 1-3, for a given $\tau\in(0,1)$, $\hat\vartheta(\tau)=(\hat\rho(\tau),\hat\beta^\top(\tau),\hat\Theta^\top(u,\tau))^\top$ converges to a Gaussian distribution:
\begin{equation}
\sqrt{n}(\hat\vartheta(\tau)-\vartheta(\tau))\stackrel{d}{\rightarrow}N(0,J^{\top}SJ),
\end{equation}
where $S=\underset{n\rightarrow\infty}{\lim}\frac{\tau(1-\tau)}{n}\tilde{X}^{\top}\tilde{X}$,
$\tilde{X}=[X,\Pi,E]$, $J=(K^{\top},L^{\top}_1,L^{\top}_2)$, $\Omega=diag(f_{i}(\xi_{i}(\tau)))$, $\mathbf{J}_{\rho}=\underset{n\rightarrow\infty}{\lim}\frac{1}{n}\tilde{X}^{\top}\Omega D$,
$\mathbf{J}_{\eta}=\underset{n\rightarrow\infty}{\lim}\frac{1}{n}\tilde{X}^{\top}\Omega \tilde{X}$, $L_1=\bar{J}^\top_{\beta}M$, $L_2=\bar{J}^\top_{\Theta}M$, $M=I-\mathbf{J}_{\rho}K$, $K=(\mathbf{J}_{\rho}^{\top}H\mathbf{J}_{\rho})^{-1}\mathbf{J}_{\rho}^{\top}H$, $H=\bar{\mathbf{J}}^{\top}_\zeta A\bar{\mathbf{J}}_\zeta$ and
$[\bar{\mathbf{J}}_{\beta},\bar{\mathbf{J}}_{\Theta},\bar{\mathbf{J}}_\zeta]$ is a
conformable partition of $\mathbf{J}^{-1}_{\eta}$.

(ii) Consequently, under Assumptions 1-3, for a given $\tau\in(0,1)$, $\hat\gamma_l(u,\tau)$, $l=1,\cdots,q$ converges to a Gaussian distribution:
\begin{equation}
\sqrt{n}(\hat\gamma_l(u,\tau)-\gamma_l(u,\tau))\stackrel{d}{\rightarrow}N(0,L_3^{(l)}SL_3^{(l)\top}),
\end{equation}
where $L_3^{(l)}=\Pi^{(l)}L_2^{(l)}$, $\Pi^{(l)}=[\Pi^{(l)\top}_1,\cdots,\Pi^{(l)\top}_n]^\top$, $\Pi^{(l)}_i=Z_{il}\pi_{k_{n}}^{\top}(U_i)$, $L_2$ is divided as $L_2=[L_2^{(1)\top},\cdots,L_2^{(q)\top}]^\top$.
\end{thm}

The confidence intervals for the coefficients are considered, which are given in the following Theorem.

\begin{thm}[Confidence Interval] \label{th:5}
(i) Under Assumptions 1-3, for a given $\tau\in(0,1)$, a $100(1-\alpha)\%$ confidence interval for the constant coefficient $\beta(\tau)$ is
$$
[\hat\beta(\tau)-\frac{Z_{\alpha/2}}{n}\sigma_\beta, \hat\beta(\tau)+\frac{Z_{\alpha/2}}{n}\sigma_\beta].
$$
where $\sigma_\beta=(\Lambda^{1/2}_{\beta11},\cdots,\Lambda^{1/2}_{\beta pp})^\top$, $\Lambda_{\beta ii}$ is the $i$th diagonal element of $\Lambda_\beta$, $\Lambda_\beta=L_1SL_1$.

(ii) Under Assumptions 1-3, for a given $\tau\in(0,1)$ and $u\in\mathcal{U}$, a $100(1-\alpha)\%$ confidence interval for the varying coefficient $\gamma_l(u,\tau)$, $l=1,\cdots,q$ is
$$
[\hat\gamma_l(u,\tau)-\frac{Z_{\alpha/2}}{n}\sigma^{(l)}_\gamma, \hat\gamma_l(u,\tau)+\frac{Z_{\alpha/2}}{n}\sigma^{(l)}_\gamma],
$$
where $\sigma^{(l)}_\gamma=(\Lambda^{(l)1/2}_{\gamma11},\cdots,\Lambda^{(l)1/2}_{\gamma nn})^\top$, $\Lambda_{\gamma ii}^{(l)}$ is the $i$th diagonal element of $\Lambda_\gamma^{(l)}$, $\Lambda_\gamma^{(l)}=L_3^{(l)}SL_3^{(l)\top}$.
\end{thm}

\section{Rank score test} \label{sec:test}

\subsection{Inference on nonvarying coefficients}

In this section, we propose a large sample inference procedures for testing the nonvarying coefficients $\beta$. We partition the original model as
\begin{align} \label{mod:06}
Q_{\tau}(y|X,Z,U)&=\rho(\tau)D+X_1\beta_1(\tau)+X_2\beta_2(\tau)+Z\gamma(u,\tau), \\
&\approx\rho(\tau)D+X_1\beta_1(\tau)+X_2\beta_2(\tau)+\Pi\Theta(\tau), \\
&=X_1\beta_1(\tau)+X^*\phi(\tau),
\end{align}
where $\beta$ are partitioned into two parts $\beta_{1}\in\mathbb{R}^{p_{1}}$ and $\beta_{2}\in\mathbb{R}^{p_{2}}$ with $p_{1}+p_{2}=p$, $X_1$ and $X_2$ are respectively $n\times p_{1}$
and $n\times p_{2}$ design matrices corresponding to $\beta_{1}$ and $\beta_{2}$, $X^*=(D,X_2,\Pi)$, $\phi=(\rho,\beta_2^{\top},\Theta^\top)^\top$.

Suppose we want to test $H_0:\beta_{1}(\tau)=0$, the quantile rank score test can be employed (see, Gutenbrunner, et al., 1990). Denote $\hat\phi(\tau)$ be the IVQR estimates of $\phi(\tau)$ obtained under $H_0$. The rank score test statistic takes the
form:
\begin{equation} \label{t:01}
RS_{n}=S_n^\top Q_n^{-1}S_n,
\end{equation}
where $S_n=n^{-\frac{1}{2}}\sum_{i=1}^{n}G_i\psi_\tau(\hat{\varepsilon}_i)$, $Q_n=n^{-1}\sum_{i=1}^{n}G_i\psi_\tau^2(\hat{\varepsilon}_i)G_i^\top$, $G=(I-P)X_1=[G_1,\cdots,G_n]^\top$, $\psi_\tau(\hat\varepsilon)=(\psi_\tau(\hat{\varepsilon}_1),\cdots,\psi_\tau(\hat{\varepsilon}_n))^\top$, $P=B^{\frac{1}{2}}X^*(X^{*\top}BX^*)^{-1}X^{*\top}B^{\frac{1}{2}}$, $B=diag(\hat{f}_1(0),\cdots,\hat{f}_n(0))$, $\hat\varepsilon(\tau)=y-X^*\hat\phi(\tau)$.

We modify Assumption 2(i) as Assumption 2(i)$^*$ and add an Assumption 4 for deriving the asymptotic distribution of the rank score statistic $RS_n$:

\textbf{Assumption 2}(i)$^*$ There exists some $r>2$ such that $\gamma_l(u)\in\mathcal{H}_r$, $l=1,\cdots,q$.

\textbf{Assumption 4} The minimum eigenvalue of $Q_n$ is bounded away from zero for sufficient large $n$.

\begin{thm} \label{th:3}
Under Assumptions 1-4 and Assumption 2(i)$^*$, suppose $n^{1/(4r)}\ll k_n\ll n^{1/4}$, then $RS_n$ has an asymptotic $\chi^{2}(p_1)$ distribution under the null hypothesis $H_{0}$.
\end{thm}

\subsection{Constancy of varying coefficients}

In this section, we also employ the rank score test for testing whether one or some of the varying coefficients is constant. Without loss of generality, we consider testing whether the first $1\leq q_1\leq q$ coefficients functions $\gamma_l(\cdot)$ are constant:
$$
H_0:\gamma_l(\tau,u)=\gamma_l(\tau), \ \ l=1,\cdots,q_1,
$$
For this purpose, we may consider the quantile regression under null hypothesis
\begin{align}\label{mod:07}
Q_{\tau}(y|X,Z,U)&=\rho(\tau)D+X\beta(\tau)+Z^*_1\gamma^*_1(\tau)+Z^*_{2}\gamma^*_2(u,\tau), \nonumber\\
&\approx\rho(\tau)D+X\beta(\tau)+Z^*_1\gamma^*_1(\tau)+\Pi_{2}\Theta_2(\tau), \nonumber\\
&=\breve{X}\varphi(\tau)+Z^*_1\gamma^*_1(\tau),
\end{align}
where $\gamma$ are partitioned into two parts $\gamma_{1}^*\in\mathbb{R}^{q_{1}}$ and $\gamma_{2}(u)^*\in\mathbb{R}^{nq_{2}}$ with $q_{1}+q_{2}=q$, $Z^*_1$ and $Z^*_2$ are respectively $n\times q_{1}$
and $n\times nq_{2}$ design matrices corresponding to $\gamma_{1}^*$ and $\gamma_{2}^*$, $\Pi_{2i}=(Z_{iq_{1}+1}\bar{\pi}_{k_n}(U_i)^\top,\cdots,Z_{iq}\bar{\pi}_{k_n}(U_i)^\top)^\top$, $\breve{X}=(D,X,\Pi_2)$, $\varphi=(\rho,\beta,\Theta_2)$.

Then we propose the test procedure as follows:
\begin{itemize}
  \item Step 1: Obtain the IVQR estimation of $\hat\gamma_1^*(\tau)$ under model \eqref{mod:07} (i.e., null hypothesis $H_0$).
  \item Step 2: We can estimate the varying coefficients $\gamma_2^*(u,\tau)$ by considering quantile regression of $y-Z_1^*\hat\gamma^*_1(\tau)$ on $\breve{X}$.
  \item Step 3: The quantile rank score test can be employed (see, Gutenbrunner, et al., 1990). Denote $\hat\varphi(\tau)$ be the IVQR estimates of $\varphi$ obtained under $H_0$. Then the rank score test statistic takes the form:
\begin{equation} \label{t:02}
RS^*_{n}=S_n^{*\top}Q_n^{*-1}S^*_n,
\end{equation}
where $S^*_n=n^{-\frac{1}{2}}\sum_{i=1}^{n}G^{*}_i\psi_\tau(\hat{\varepsilon}_i)$, $Q^*_n=n^{-1}\sum_{i=1}^{n}G^{*}_i\psi_\tau^2(\hat{\varepsilon}_i)G^{*\top}_i$, $\psi_\tau(\hat\varepsilon)=(\psi_\tau(\hat{\varepsilon}_1),\cdots,\psi_\tau(\hat{\varepsilon}_n))^\top$, $G^*=(I-P^*)Z_1^*=[G_1^*,\cdots,G_n^*]^\top$, $P^*=B^{\frac{1}{2}}\breve{X}(\breve{X}^{\top}B\breve{X})^{-1}\breve{X}^{\top}B^{\frac{1}{2}}$, $B=diag(\hat{f}_1(0),\cdots,\hat{f}_n(0))$, $\hat\varepsilon=y-Z_1^*\hat\gamma^*_1(\tau)-\breve{X}\hat{\varphi}(\tau)$.
\end{itemize}

We modify Assumption 4 as Assumption 4$^*$ for deriving the asymptotic distribution of the rank score statistic $RS_n^*$:

\textbf{Assumption 4}$^*$ The minimum eigenvalue of $Q^*_n$ is bounded away from zero for sufficient large $n$.

\begin{thm} \label{th:4}
(i) If $k_n=k$ is bounded corresponding to model \eqref{mod:07}, then under Assumptions 1-3, Assumption 2(i)$^*$ and Assumption 4$^*$, suppose $n^{1/(4r)}\ll k_n\ll n^{1/4}$, then $RS_n^*$ has an asymptotic $\chi^{2}(q_1)$ distribution under the null hypothesis $H_{0}$.

(ii) For growing $k_n$ as the sample size $n$ becomes larger, then under Assumptions 1-3, Assumption 2(i)$^*$ and Assumption 4$^*$, $h\geq3$, suppose the number of knots satisfies $n^{1/(2r+2)}\ll k_n\ll n^{1/5}$, then under $H^*_0$, we have
\begin{equation}
\frac{RS_n^*-q_1}{\sqrt{2q_1}}\stackrel{d}{\rightarrow}N(0,1) \ \ as\ k_n\rightarrow\infty.
\end{equation}
\end{thm}

\section{Monte Carlo simulations} \label{sec:simu}

In this section, we conduct Monte Carlo simulations to investigate the finite sample performance of the proposed estimation and inference methods. The Monte Carlo simulations are repeated 1000 times for each sample size $n=100, 200, 500, 800$. The quantile regression based estimators are calculated for quantiles $\tau=(0.25,0.5,0.75)$.

\textit{Example 1}. The samples are generated as follows:
\begin{equation} \label{eg:1}
y_i=\rho\sum_{i=1}^{n}w_{ij}y_j+X_i\beta+Z_{1i}\gamma_1(U_i)+Z_{2i}\gamma_2(U_i)+\varepsilon_i, \ \ i=1,\cdots,n,
\end{equation}
where $\rho=0.5$, $\beta=1$, $\gamma_1(U)=1-0.5U$, $\gamma_2(U)=1+\sin(\sqrt{2}\pi U)$, $\varepsilon_{i}=e_{i}-F^{-1}(\tau)$, $F$ is the common CDF of $e_{i}$. Therefore, the random errors $\varepsilon_i$ are centered to have zero $\tau$th quantile. Here, $U,X,Z_1,Z_2,e$ respectively follow the $U[0,2]$, $N(0,1)$, $U[-2,2]$, $N(1,1)$ and $N(0,1)$ distributions.

\textit{Example 2}. The samples are generated as follows:
\begin{equation}\label{eg:2}
y_i=\rho\sum_{i=1}^{n}w_{ij}y_j+X_i\beta+Z_{1i}\gamma_1(U_i)+Z_{2i}\gamma_2(U_i)+(1+0.5Z_{1i})\varepsilon_i, \ \ i=1,\cdots,n,
\end{equation}
where $\rho=0.5$, $\beta=1$, $\gamma_1(U)=1-0.5U$, $\gamma_2(U)=0.5U^2-U+1$, $\varepsilon_{i}=e_{i}-F^{-1}(\tau)$, $F$ is the common CDF of $e_{i}$. Therefore, the random errors $\varepsilon_i$ are centered to have zero $\tau$th quantile. In this example, $U,X,Z_1,Z_2,e$ respectively follow the $U[0,2]$, $N(0,1)$, $N(0,1)$, $U[-2,2]$ and $N(0,1)$ distributions.

Following Dai, et al. (2016), the spatial weight
matrix $W=(w_{ij})$ in the two examples is generated based on mechanism that $w_{ij} =r^{|i-j|}I(i\neq j)$, where $r=0.3$, $0<i,j<n$. A standardized transformation then is used to convert the matrix $W$ to
have row-sums of unit.

\subsection{Estimation}

Firstly, we compare the performance of the partially linear varying coefficient spatial autoregressive model to the spatial autoregressive model. In example 1, the spatial autoregressive model is of the form
\begin{equation} \label{eg:11}
y_i=\rho\sum_{i=1}^{n}w_{ij}y_j+X_i\beta+Z_{1i}\gamma_1+Z_{2i}\gamma_2+\varepsilon_i, \ \ i=1,\cdots,n,
\end{equation}
where $\gamma_1=1$, $\gamma_2=1$, the rest variables are the same as those defined in model \eqref{eg:1}. In example 2, the spatial autoregressive model is given by
\begin{equation}\label{eg:21}
y_i=\rho\sum_{i=1}^{n}w_{ij}y_j+X_i\beta+Z_{1i}\gamma_1+Z_{2i}\gamma_2+(1+0.5Z_i)\varepsilon_i, \ \ i=1,\cdots,n,
\end{equation}
where $\gamma_1=1$, $\gamma_2=1$, the rest variables are the same as those defined in model \eqref{eg:2}. Table 1 gives the comparison results of bias and RMSE of the PLVCSAR model and SAR model at $\tau=\{0.25, 0.5, 0.75\}$ and $n=100$. $\hat\rho_{PLVC}$ and $\hat\beta_{PLVC}$ denote the IVQR estimates in PLVCSAR models, and $\hat\rho_{SAR}$ and $\hat\beta_{SAR}$ denote the IVQR estimates in SAR models. From Table 1, we can see that when data is generated from the PLVCSAR model, fitting SAR model leads to less efficient estimations in two examples, the bias and RMSE of $\hat\rho_{PLVC}$ and $\hat\beta_{PLVC}$ is smaller than those of $\hat\rho_{SAR}$ and $\hat\beta_{SAR}$. When data is generated from the SAR model, fitting PLVCSAR model and SAR model have similar performance in homoscedastic case; in heteroscedastic case, fitting PLVCSAR model still does not lose much efficiency. Thus the PLVCSAR model is efficient and more flexible than the SAR model.

Table 2 summarizes the comparison results of QR and IVQR estimators with homoscedastic error term. Table 3 reports the comparison results of QR and IVQR estimators with heteroscedastic error term. Table 2 and 3 show that the IVQR estimator of $\rho$ has much smaller bias and RMSE than QR estimator on the whole, and the IVQR estimators of $\beta$ and $\gamma$ have similar bias and RMSE as QR estimators.

The confidence intervals of the varying coefficients are also considered. The results are reported in Figure 1. The $x$-axis presents the smoothing variables, and $y$-axis presents the estimations of the varying coefficients at quantile 0.5 and sample size 200 (red lines) and their corresponding confidence intervals (blue lines) at significance level 0.05. Figure 1(a)-(b) and (c)-(d) respectively gives the confidence intervals of $\gamma_1, \gamma_2$ in Example 1 (with homoscedastic error term) and Example 2 (with heteroscedastic error term).

\renewcommand{\baselinestretch}{1.0}
\begin{table}[t]\footnotesize
\caption{Comparison results of SAR and PLVCSAR models. The table shows the bias and RMSE (in parentheses) for $\rho,\beta$ at $\tau=\{0.25,0.5,0.75\}$ quantile and $n=100$. }
\begin{center}
\begin{tabular}{ccccccccc} \toprule
\multirow{2}*{Example}&\multirow{2}*{Parameter}&\multicolumn{3}{c}{Underlying model: PLVCSAR}&
&\multicolumn{3}{c}{Underlying model: SAR} \\ \cline{3-5} \cline{7-9}
        &   &$\tau=0.25$&$\tau=0.50$&$\tau=0.75$& &$\tau=0.25$&$\tau=0.50$&$\tau=0.75$ \\ \hline
1&$\hat\rho_{PLVC}$&0.0021&0.0065&0.0077&&-0.0012&0.0042&0.0037   \\
& &(0.1302)&(0.1246)&(0.1311)&&(0.1144)&(0.1011)&(0.1133)   \\
&$\hat\rho_{SAR}$&0.0089&0.0158&0.0239&&0.0046&0.0048&0.0035   \\
& &(0.1777)&(0.1577)&(0.1562)&&(0.1119)&(0.1014)&(0.1132)   \\
&$\hat\beta_{PLVC}$&-0.0021&-0.0006&-0.0086&&-0.0010&0.0002&-0.0062   \\
& &(0.1516)&(0.1408)&(0.1523)&&(0.1524)&(0.1334)&(0.1464)   \\
&$\hat\beta_{SAR}$&-0.0139&-0.0060&-0.0093&&-0.0044&-0.0039&-0.0087   \\
& &(0.2032)&(0.1854)&(0.1842)&&(0.1423)&(0.1305)&(0.1463)   \\
& & & & & & & & \\
2&$\hat\rho_{PLVC}$&0.0070&0.0011&0.0009&&0.0069&0.0037&0.0044   \\
& &(0.1289)&(0.1197)&(0.1289)&&(0.0943)&(0.0973)&(0.1099)   \\
&$\hat\rho_{SAR}$&0.0074&0.0080&0.0398&&0.0015&0.0067&0.0016   \\
& &(0.1340)&(0.1383)&(0.1630)&&(0.0855)&(0.0814)&(0.1042)   \\
&$\hat\beta_{PLVC}$&-0.0074&0.0014&-0.0026&&0.0021&-0.0031&-0.0022   \\
& &(0.1326)&(0.1155)&(0.1325)&&(0.1259)&(0.1211)&(0.1238)   \\
&$\hat\beta_{SAR}$&-0.0078&-0.0081&-0.0047&&-0.0003&-0.0020&-0.0077   \\
& &(0.1365)&(0.1225)&(0.1447)&&(0.1124)&(0.1038)&(0.1136)   \\
 \bottomrule
\end{tabular}
\end{center}
\end{table}
\renewcommand{\baselinestretch}{1.5}

\renewcommand{\baselinestretch}{1.0}
\begin{table}[t]\footnotesize
\caption{Monte Carlo results for $\tau=\{0.25,0.5,0.75\}$ quantile and homoscedastic error term. The table shows the bias and RMSE (in parentheses) for $\rho,\beta$, and the MADE [in brackets] for $\gamma_1,\gamma_2$. }
\begin{center}
\begin{tabular}{lcccccccc} \toprule
\multirow{2}*{Sample size}& &\multicolumn{3}{c}{QR}&
&\multicolumn{3}{c}{IVQR} \\ \cline{3-5} \cline{7-9}
        &   &$\tau=0.25$&$\tau=0.50$&$\tau=0.75$& &$\tau=0.25$&$\tau=0.50$&$\tau=0.75$ \\ \hline
$n=100$&$\rho$&0.0214&0.0373&0.0528&&0.0037&0.0025&0.0021   \\
& &(0.0516)&(0.0700)&(0.0993)&&(0.1315)&(0.1186)&(0.1329)   \\
&$\beta$&-0.0063&-0.0036&-0.0149&&-0.0065&-0.0030&0.0041    \\
& &(0.1440)&(0.1334)&(0.1460)&&(0.1431)&(0.1364)&(0.1508)   \\
&$\gamma_1$&[0.2203]&[0.1973]&[0.2207]&&[0.2202]&[0.2031]&[0.2200]   \\
&$\gamma_2$&[0.2038]&[0.1930]&[0.2002]&&[0.2139]&[0.1971]&[0.2145]   \\
$n=200$&$\rho$&0.0198&0.0341&0.0569&&0.0016&0.0008&-0.0011   \\
& &(0.0372)&(0.0527)&(0.0804)&&(0.0853)&(0.0761)&(0.0859)   \\
&$\beta$&-0.0054&-0.0044&-0.0171&&0.0003&-0.0016&0.0021   \\
& &(0.1010)&(0.0930)&(0.1035)&&(0.1009)&(0.0918)&(0.0966)   \\
&$\gamma_1$&[0.1479]&[0.1379]&[0.1491]&&[0.1520]&[0.1377]&[0.1515]   \\
&$\gamma_2$&[0.1533]&[0.1425]&[0.1452]&&[0.1513]&[0.1423]&[0.1530]   \\
$n=500$&$\rho$&0.0213&0.0384&0.0572&&0.0025&-0.0006&0.0009   \\
& &(0.0297)&(0.0463)&(0.0672)&&(0.0539)&(0.0462)&(0.0520)   \\
&$\beta$&-0.0008&-0.0103&-0.0106&&-0.0011&-0.0002&-0.0010   \\
& &(0.0600)&(0.0590)&(0.0635)&&(0.0599)&(0.0600)&(0.0635)   \\
&$\gamma_1$&[0.0925]&[0.0862]&[0.0921]&&[0.0919]&[0.0857]&[0.0914]   \\
&$\gamma_2$&[0.1066]&[0.1083]&[0.1044]&&[0.1040]&[0.1027]&[0.1041]   \\
$n=800$&$\rho$&0.0226&0.0362&0.0599&&-0.0002&-0.0006&-0.0004   \\
& &(0.0280)&(0.0413)&(0.0660)&&(0.0405)&(0.0385)&(0.0402)   \\
&$\beta$&-0.0038&-0.0064&-0.0116&&-0.0029&-0.0020&0.0005   \\
& &(0.0486)&(0.0451)&(0.0485)&&(0.0478)&(0.0443)&(0.0476)   \\
&$\gamma_1$&[0.0721]&[0.0675]&[0.0722]&&[0.0711]&[0.0674]&[0.0701]   \\
&$\gamma_2$&[0.0981]&[0.0956]&[0.0947]&&[0.0741]&[0.0892]&[0.0924]   \\ \bottomrule
\end{tabular}
\end{center}
\end{table}
\renewcommand{\baselinestretch}{1.5}

\renewcommand{\baselinestretch}{1.0}
\begin{table}[t] \footnotesize
\caption{Monte Carlo results for $\tau=\{0.25,0.5,0.75\}$ quantile and heteroscedastic error term. The table shows the bias and RMSE (in parentheses) for $\rho,\beta$, and the MADE [in brackets] for $\gamma_1,\gamma_2$. }
\begin{center}
\begin{tabular}{lcccccccc} \toprule
\multirow{2}*{}& &\multicolumn{3}{c}{QR}&
&\multicolumn{3}{c}{IVQR} \\ \cline{3-5} \cline{7-9}
        &   &$\tau=0.25$&$\tau=0.50$&$\tau=0.75$& &$\tau=0.25$&$\tau=0.50$&$\tau=0.75$ \\ \hline
$n=100$&$\rho$&0.0477&0.0861&0.0560& &0.0070&0.0011&0.0009  \\
& &(0.0835)&(0.1252)&(0.0915)& &(0.1289)&(0.1197)&(0.1289)  \\
&$\beta$ &-0.0147&-0.0204&-0.0101& &-0.0074&0.0014&-0.0026  \\
& &(0.1298)&(0.1222)&(0.1309)& &(0.1326)&(0.1155)&(0.1325)  \\
&$\gamma_1$ &[0.2257]&[0.1892]&[0.2323]& &[0.2317]&[0.1989]&[0.2405]  \\
&$\gamma_2$ &[0.1953]&[0.1782]&[0.1982]& &[0.2004]&[0.1775]&[0.2030]  \\
$n=200$&$\rho$&0.0445&0.0874&0.0531& &-0.0004&0.0036&-0.0007  \\
& &(0.0638)&(0.1049)&(0.0723)& &(0.0801)&(0.0740)&(0.0907) \\
&$\beta$ &-0.0114&-0.0177&-0.0133& &-0.0029&0.0008&-0.0056  \\
& &(0.0837)&(0.0786)&(0.0827)& &(0.0819)&(0.0736)&(0.0839)  \\
&$\gamma_1$ &[0.1337]&[0.1093]&[0.1421]& &[0.1406]&[0.1139]&[0.1403]  \\
&$\gamma_2$ &[0.1231]&[0.1141]&[0.1244]& &[0.1256]&[0.1117]&[0.1235]  \\
$n=500$&$\rho$&0.0433&0.0804&0.0510& &-0.0013&0.0009&-0.0001  \\
& &(0.0512)&(0.0878)&(0.0585)& &(0.0440)&(0.0429)&(0.0517)   \\
&$\beta$ &-0.0104&-0.0140&-0.0117& &0.0008&0.0008&-0.0001  \\
& &(0.0466)&(0.0464)&(0.0473)& &(0.0476)&(0.0420)&(0.0484)  \\
&$\gamma_1$ &[0.0789]&[0.0548]&[0.0906]& &[0.0757]&[0.0582]&[0.0766]   \\
&$\gamma_2$ &[0.0703]&[0.0643]&[0.0691]& &[0.0706]&[0.0628]&[0.0715]  \\
$n=800$&$\rho$&0.0417&0.0776&0.0495& &0.0013&0.0013&-0.0000  \\
& &(0.0462)&(0.0823)&(0.0545)& &(0.0326)&(0.0340)&(0.0380) \\
&$\beta$ &-0.0083&-0.0174&-0.0081& &-0.0042&-0.0009&-0.0014  \\
& &(0.0367)&(0.0381)&(0.0347)& &(0.0358)&(0.0317)&(0.0346)  \\
&$\gamma_1$ &[0.0657]&[0.0408]&[0.0762]& &[0.0582]&[0.0402]&[0.0585] \\
&$\gamma_2$ &[0.0522]&[0.0502]&[0.0530]& &[0.0521]&[0.0483]&[0.0526]  \\ \bottomrule
\end{tabular}
\end{center}
\end{table}
\renewcommand{\baselinestretch}{1.5}

\begin{figure}[t]
\begin{center}
\scalebox{0.8}[0.8]{\includegraphics{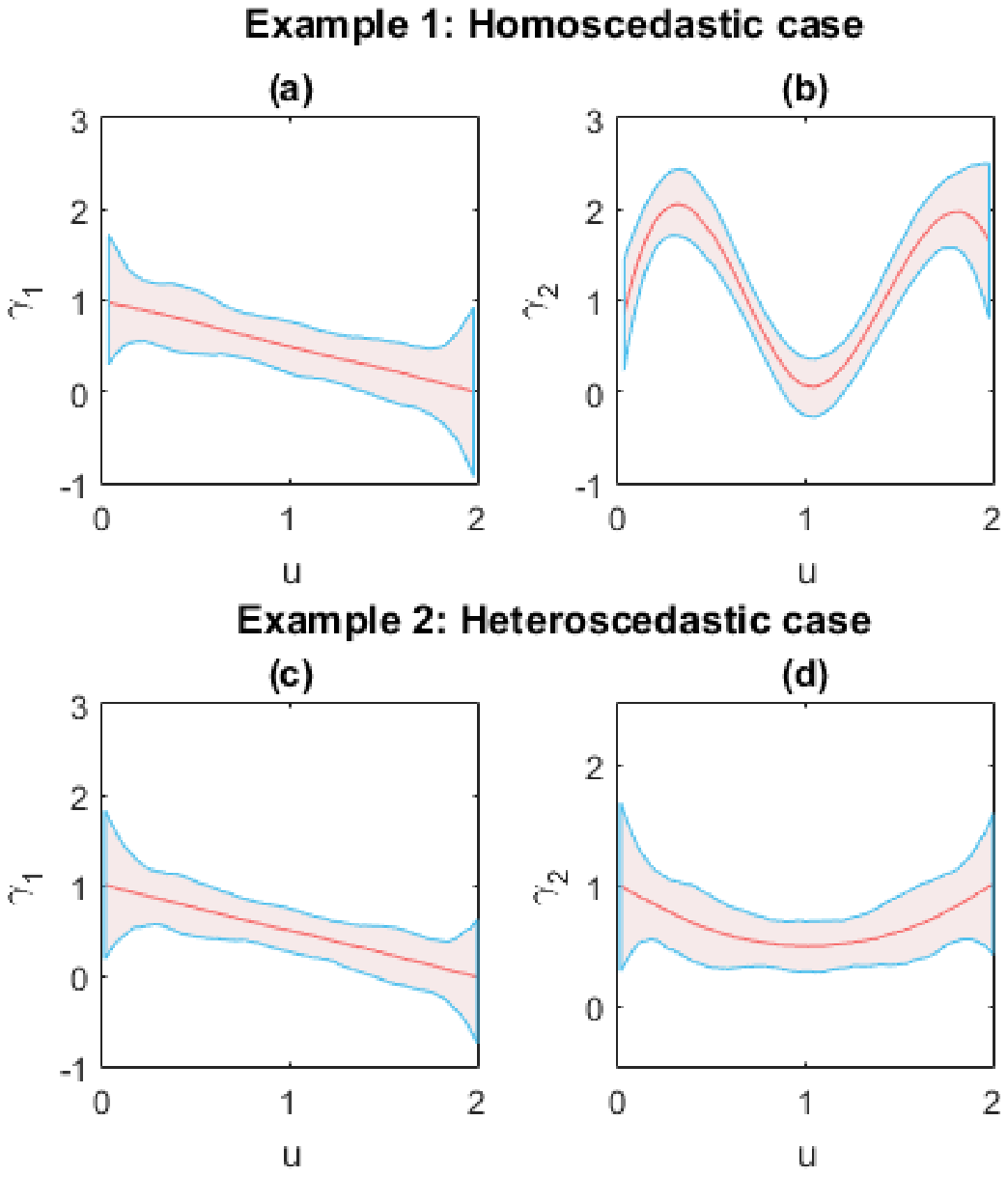}}
\end{center}
\caption{(a)-(b) Confidence intervals of $\gamma_1$ and $\gamma_2$ in Example 1 (homoscedastic case) at $\tau=0.5$ and $n=200$. (c)-(d) Confidence intervals of $\gamma_1$ and $\gamma_2$ in Example 2 (heteroscedastic case) at $\tau=0.5$ and $n=200$. The areas represent 95\% point-wise confidence intervals. }
\end{figure}

\subsection{Inference on $\beta$}

To study the size and power of test statistics $RS_{n}$ and $RS_{n}$, we vary $\beta$ in model \eqref{eg:1} and \eqref{eg:2} from 0 to 1.5. The result is listed in the left 5 columns of Table 4, from which we can see that the size of test statistics $RS_{n}$ is much close to the nominal significant level 0.05 compared with test statistics $RS_{n}$. The power of test statistics $RS_{n}$ and $RS_{n}$ are not clearly different.

\renewcommand{\baselinestretch}{1.0}
\begin{table}[t]\footnotesize
\begin{center}
\caption{Size and power for testing $H_0:\beta=0$ and $H_0:\gamma_1(u)=\gamma_1$ at $\tau=0.5$ and $n=200$. Here the nominal significance level is 0.05. }
\begin{tabular}{cccccccccccc} \toprule
&\multicolumn{5}{c}{$H_0:\beta=0$}&&\multicolumn{5}{c}{$H_0:\gamma_1(u)=\gamma_1$} \\ \cline{2-6}\cline{8-12}
&\multicolumn{2}{c}{Example 1}&&\multicolumn{2}{c}{Example 2}&&\multicolumn{2}{c}{Example 1}&&\multicolumn{2}{c}{Example 2} \\ \cline{2-3}\cline{5-6}\cline{8-10}\cline{11-12}
$\beta$&QR&IVQR&&QR&IVQR&$\eta$&QR&IVQR&&QR&IVQR \\ \hline
0&0.0619&0.0530&&0.0548&0.0486&0&0.0400&0.0440&&0.0602&0.0534 \\
0.25&0.8680&0.7570&&0.4730&0.4530&0.25&0.2460&0.2710&&0.3320&0.3440\\
0.5&0.9990&0.9970&&0.8990&0.8820&0.5&0.7940&0.8242&&0.7610&0.7570 \\
0.75&1.0000&1.0000&&0.9900&0.9910&0.75&0.9920&0.9920&&0.9490&0.9430 \\
1&1.0000&1.0000&&0.9998&1.0000&1&0.9970&1.0000&&0.9930&0.9840 \\
1.25&1.0000&1.0000&&1.0000&1.0000&1.25&1.0000&1.0000&&0.9996&1.0000 \\
1.5&1.0000&1.0000&&1.0000&1.0000&1.5&1.0000&1.0000&&1.0000&1.0000 \\
 \bottomrule
\end{tabular}
\end{center}
\end{table}
\renewcommand{\baselinestretch}{1.5}

\subsection{Inference on the constancy of $\gamma(U)$}

To test whether $\gamma_1(U)$ is constant, we generate $\gamma_1(U)$ from
$$
\gamma_1(U)=1-0.5*\eta*U,
$$
where $\eta$ varies from 0 to 1.5. The results of size and power of test statistics $RS^*_{n}$ and $RS_{n}^{*IV}$ are reported in the right 5 columns of Table 4. Table 4 shows that the size of test statistics $RS_{n}^{*IV}$ is much close to the nominal significant level 0.05 compared with test statistics $RS_{n}^*$. The power of test statistics $RS_{n}^*$ and $RS_{n}^{*IV}$ are not clearly different.

\section{Illustration} \label{sec:real}

In this section, we apply the proposed estimation method to the Boston housing price data, which has been analyzed by many authors (see, LeSage, 2009; Tang, et al., 2013; Sun, et al., 2014; Dai, et al., 2016).
The data set contains 14 variables with 506 observations. The latitude and longitude coordinates
are also provided. Sun, et al. (2014) chose the following five variables as explanatory variables, such as the per capita crime rate by town
($X_1$), average number of rooms per dwelling ($X_2$), index of accessibility to radial highways ($X_3$),
full-value property-tax rate per \$10,000 dollar ($X_4$) and the percentage of the lower status of the population ($X_5$), and the dependent variable $y$ is the median value of owner-occupied homes
in \$1000s. They employed a partially linear varying coefficient spatial autoregressive model (SAR) to analyze this data set in the mean regression framework, and applied BIC coupled with backward elimination to do the model selection and identify the constancy of the coefficients.

In this section, we first employ the proposed method for testing the constancy of the coefficients. We consider a set of quantiles with $\tau=\{0.1, 0.2,\cdots,0.9\}$. At each quantile $\tau$, We consider five null hypothesis, $H_{0i}$: $\gamma_i(u,\tau)$ is constant, $i=1,\cdots,5$. The results are summarized in Table 5. From Table 5, we can see that the effect of $X_2$ is varying at quantile 0.2-0.9, the effect of $X_3$ is varying at quantile 0.6, the effect of $X_5$ is varying at quantile 0.4-0.8, and the coefficients of the other two variables are constant at all quantile levels.

It is noted that the result is a little different from Sun, et al. (2014). In which, coefficients of $X_1$, $X_2$ and $X_4$ are chosen as varying-coefficient. To demonstrate our results, we compare the SIC values in the following four models: PLVCSAR model1 (with varying coefficients $X_2$ and $X_5$), PLVCSAR model2 (with varying coefficients $X_2$, $X_3$ and $X_5$), PLVCSAR model3 (with varying coefficients $X_1$, $X_2$ and $X_4$), and constant coefficient SAR model. The comparison results are summarized in Table 6. Generally speaking, PLVCSAR model1 (with varying coefficients $X_2$ and $X_5$) has the smallest SIC values. Thus the following model is considered:
\begin{align}\label{mod:31}
Q_\tau(y_i|\mathcal{F}_{-i},X_{1i},\cdots,X_{5i})&=\rho(\tau)\sum_{j=1}^{n}w_{ij}y_j+X_{i1}
\beta_1(\tau)+X_{i2}\beta_2(U_i,\tau)+X_{i3}\beta_3(\tau)\nonumber \\
&\quad+X_{i4}\beta_4(\tau)+X_{i5}\beta_5(U_i,\tau),
\end{align}
where $w_{ij}$ is the $(i,j)$th element of $W$, $W$ is the spatial weight matrix generated by the longitude and latitude of the 506 observations (LeSage, 1999, p68).

Table 7 reports the estimates of the constant coefficients $\rho,\gamma_1,\gamma_3$ and $\gamma_4$, which shows the per capita crime rate by town $(X_1)$ and full-value property-tax rate per \$10,000 dollar ($X_4$) has negative impact on house price, the index of accessibility to radial highways $(X_3)$ has positive impact on
house price. Besides, the coefficient of spatial correlation is 0.1 at $\tau=0.1, 0.3, 0.7$, 0.05 at $\tau=0.5$, and -0.25 at $\tau=0.9$, which indicates the house prices in a neighborhood do affect each other.

Figure 2(a)-(b) plot the surface of the estimated $\gamma_2$ and $\gamma_5$. The $x$-axis presents the quantiles, $y$-axis presents the smoothing variables, and $z$-axis presents the estimations of the varying coefficients. From Figure 2, we can see the impact $\gamma_2(u,\tau)$ of the average number of rooms per dwelling on house price is positive and is varying over location. The impact $\gamma_5(u,\tau)$ of the percentage of the lower status of the population on house price is also varying over location.

Figure 3(a)-(c) and Figure 4(a)-(f) present a complete analysis, which considers many other quantiles of the conditional boston house price distribution. The $x$-axis presents the quantiles and $y$-axis presents the estimations of  parameters (red lines) and their corresponding confidence intervals (blue lines) at significance level 0.05. Figure 3(a)-(c) present the results of constant coefficient $\gamma_1,\gamma_3$ and $\gamma_4$, from which we can find that the estimates of the per capita crime rate by town $(X_1)$ are smaller at the middle quantiles than those at other quantiles. The estimates of the index of accessibility to radial highways $(X_3)$ are increasing as quantile becomes larger. And the estimates of full-value property-tax rate per \$10,000 dollar ($X_4$) generally does not vary with quantiles.

Figure 4(a)-(c) present the results of varying coefficient $\gamma_2(u)$ at $u=0.5,1$ and 1.5. On the whole, the estimates of average number of rooms per dwelling ($X_2$) is ascending as quantile becomes larger. Figure 4(d)-(f) present the results of varying coefficient $\gamma_5(u)$ at $u=0.5,1$ and 1.5. From Figure 4(d)-(f), we can see that at $u=0.5$, the estimates of the percentage of the lower status of the population ($X_5$) decease at quantile 0.1-0.3, then ascend at quantile 0.3-0.65, and then descends at high quantiles. At $u=1$, $\hat\gamma_5(u,\tau)$ is larger at extreme quantiles. At $u=1.5$, $\hat\gamma_5(u,\tau)$ is descending at quantile 0.1-0.35, and then increases at quantiles 0.3-0.85 and then suddenly deceases.

\renewcommand{\baselinestretch}{1.0}
\begin{table}[t]\footnotesize
\begin{center}
\caption{Rank Score statistics and corresponding cut-off value for
identifying single varying coefficient. $H_{0l}$ represents testing the constancy of $\beta_l(u,\tau)$, $l=1,\cdots,5$. Here the nominal significance level is 0.05. Values marked with * correspond to rejection of null hypothesis.}
\begin{tabular}{cccccc} \toprule
$\tau$&$H_{01}$&$H_{02}$&$H_{03}$&$H_{04}$&$H_{05}$ \\ \hline
0.1&0.0001&1.3179&0.0332&0.2787&1.4386 \\
0.2&0.0002&10.5218 *&0.0428&0.7629&2.8423 \\
0.3&0.0001&22.1575 *&0.0013&0.9900&1.5555 \\
0.4&0.0716&5.5943 *&0.6882&0.0000 &10.2199 * \\
0.5&0.2178&9.8951 *&1.0426&0.0029&6.8899 *  \\
0.6&0.8100&13.6297 *&4.8560 *&0.0318&6.0768 * \\
0.7&0.6145&8.4320 *&0.8975&0.0923&7.3477 * \\
0.8&0.8827&4.0135 *&0.0649&0.0203&4.9928 * \\
0.9&0.0001&4.0135 *&0.0218&0.0014&1.4386  \\
&\multicolumn{5}{c}{Cut-off value : 3.8415} \\ \bottomrule
\end{tabular}
\end{center}
\end{table}
\renewcommand{\baselinestretch}{1.5}

\renewcommand{\baselinestretch}{1.0}
\begin{table}[t]\footnotesize
\begin{center}
\caption{SIC values of three models. $(X_2,X_5)$ represents the PLVCSAR model1 with varying coefficients $X_2$ and $X_5$, $(X_2,X_3,X_5)$ represents the PLVCSAR model2 with varying coefficients $X_2$, $X_3$ and $X_5$, $(X_1,X_2,X_4)$ represents the PLVCSAR model3 with varying coefficients $X_1$, $X_2$ and $X_4$, SAR model represents the constant coefficient spatial autoregressive model. }
\begin{tabular}{cccccccccc} \toprule
Models&$\tau=0.1$&$\tau=0.2$&$\tau=0.3$&$\tau=0.4$
&$\tau=0.5$&$\tau=0.6$&$\tau=0.7$&$\tau=0.8$&$\tau=0.9$ \\ \hline
$(X_2,X_5)$&10.4175&10.4659&10.4624&8.8761&8.1486&9.7401&9.7439&10.3754&8.9282   \\
$(X_2,X_3,X_5)$&11.5136&10.7574&10.2324&9.2940&8.6772&10.4068&10.9622&10.6783&11.9817   \\
$(X_1,X_2,X_4)$&11.4568&11.1219&10.3938&10.1365&9.3265&10.5849&10.3927&10.9940&11.4164   \\
SAR model&9.8621&8.1950&9.6300&10.2494&10.7568&11.4899&11.8461&12.1130&12.4325   \\
 \bottomrule
\end{tabular}
\end{center}
\end{table}
\renewcommand{\baselinestretch}{1.5}

\renewcommand{\baselinestretch}{1.0}
\begin{table}[t]\footnotesize
\begin{center}
\caption{Estimates of the constant coefficients at quantile $\tau=0.1, 0.3, 0.5, 0.7, 0.9$. }
\begin{tabular}{cccccc} \toprule
&\multicolumn{5}{c}{IVQR} \\ \cline{2-6}
&$\tau=0.1$&$\tau=0.3$&$\tau=0.5$&$\tau=0.7$&$\tau=0.9$ \\ \hline
$\rho$&0.1000&0.1000&0.0500&0.1000&-0.2500  \\
$\gamma_1$&-0.0557&-0.0658&-0.0677&-0.0722&-0.0215  \\
$\gamma_3$&0.2211&0.2297&0.2508&0.3892&0.6088   \\
$\gamma_4$&-0.0163&-0.0107&-0.0078&-0.0131&-0.0056   \\ \bottomrule
\end{tabular}
\end{center}
\end{table}
\renewcommand{\baselinestretch}{1.5}

\begin{figure}[t]
\begin{center}
\scalebox{0.8}[0.8]{\includegraphics{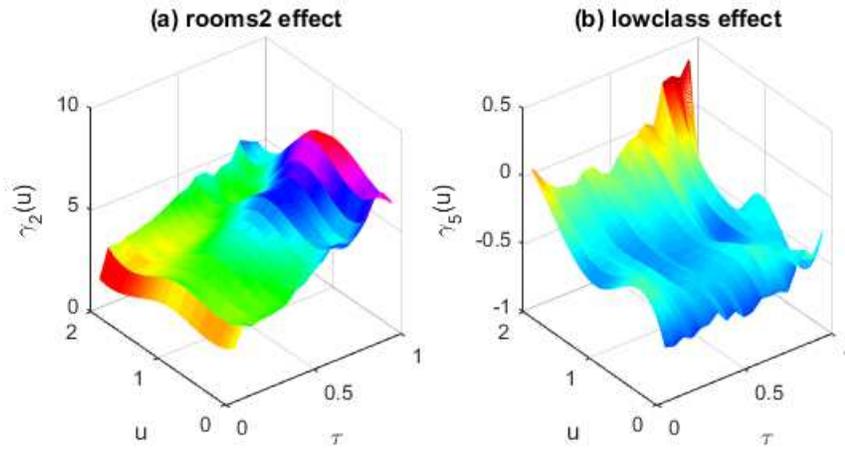}}
\end{center}
\caption{The estimated varying coefficient surface. }
\end{figure}

\begin{figure}[t]
\begin{center}
\scalebox{0.75}[0.75]{\includegraphics{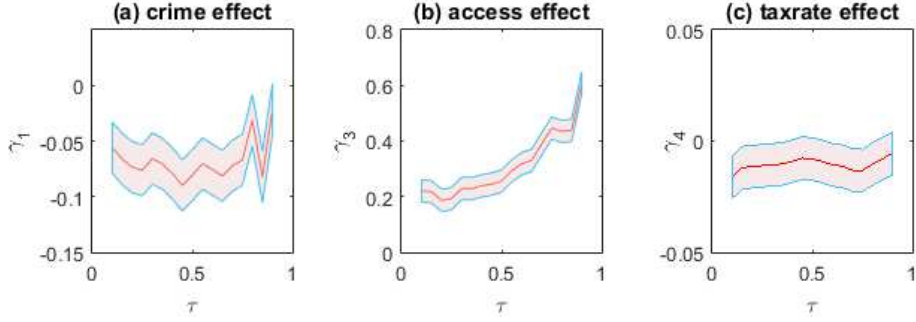}}
\end{center}
\caption{(a)-(c) Quantile effects of the per capita crime rate by town ($X_1$), index of accessibility to radial highways ($X_3$) and full-value property-tax rate per \$10,000 dollar ($X_4$). The areas represent 95\% point-wise confidence intervals. }
\end{figure}

\begin{figure}[t]
\begin{center}
\scalebox{0.75}[0.75]{\includegraphics{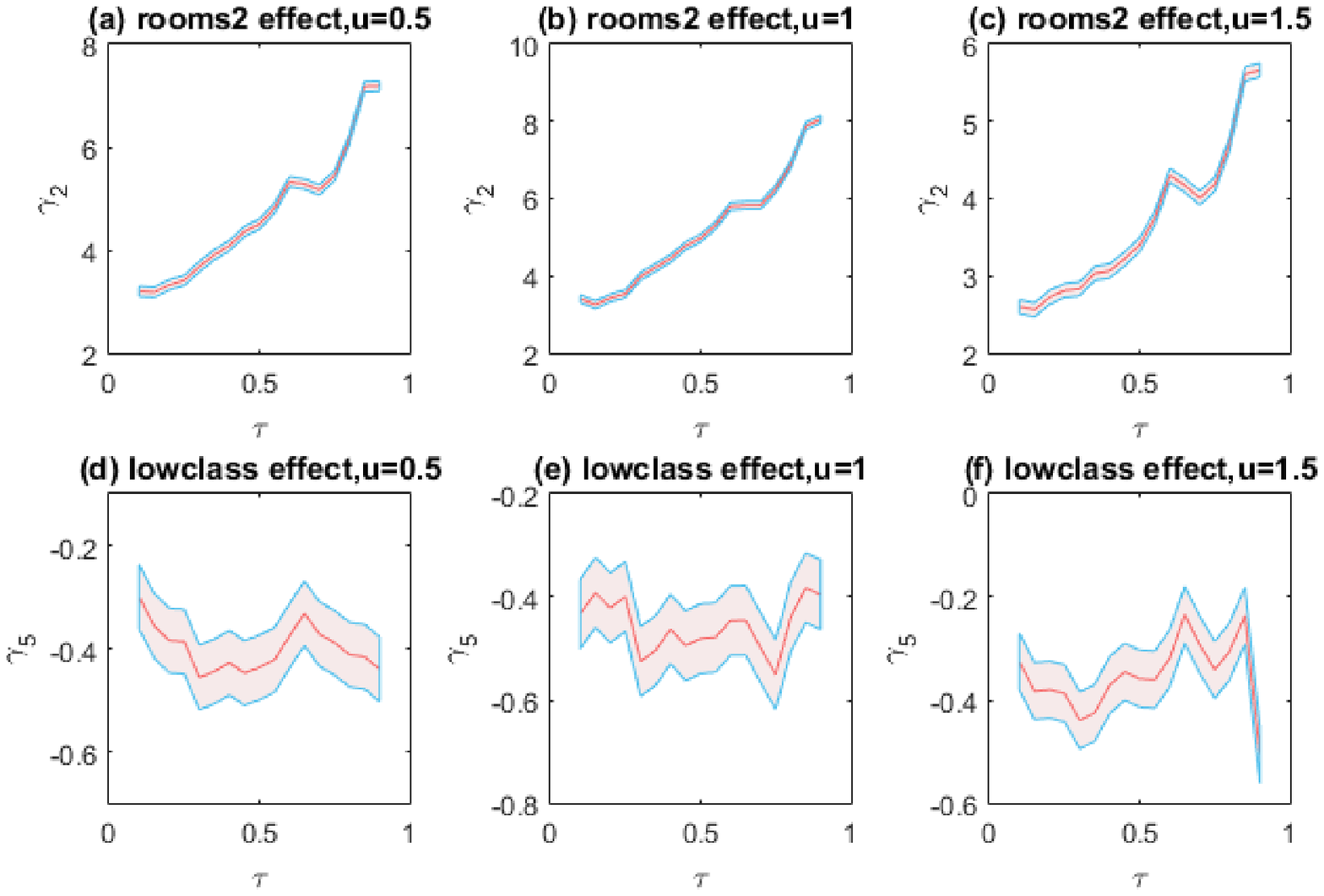}}
\end{center}
\caption{(a)-(c) Quantile effects of average number of rooms per dwelling ($X_2$) at $u=0.5,1$ and 1.5.
(d)-(f) Quantile effects of the percentage of the lower status of the population ($X_5$) at $u=0.5,1$ and 1.5. The areas represent 95\% point-wise confidence intervals. }
\end{figure}

\section{Conclusion} \label{sec:con}

In this paper, we consider IVQR estimation of partially linear varying coefficient spatial autoregressive model. The varying coefficients are approximated by B-spline basis. Rank score tests are employed for inference on $\beta$ and $\gamma(u)$. The asymptotic properties of estimators and test statistics are studied. The proposed methodology in this paper does not need any specification of error distribution. Monte Carlo results are provided to show that the IVQR estimation method can significantly reduce estimation bias. The example analysis shows the effectiveness of our estimator and test. Besides, the confidence interval of constant coefficients and varying coefficients are also given.

\section{Acknowledgements}

The work was partially supported by National Natural Science Foundation of China (No.11271368), Project supported by the Major Program of Beijing Philosophy and Social Science Foundation of China (No. 15ZDA17), Project of Ministry of Education supported by the Specialized Research Fund for the Doctoral Program of Higher Education of China (Grant No. 20130004110007), The Key Program of National Philosophy and Social Science Foundation Grant (No. 13AZD064), The Fundamental Research Funds for the Central Universities, and the Research Funds of Renmin University of China (No. 15XNL008), and The Project of Flying Apsaras Scholar of Lanzhou University of Finance \& Economics.

\section*{Appendix: Proofs}

To prove Theorems \ref{th:1} and \ref{th:2}, we first state a lemma whose proof is similar as that of Lemma 2 in Galvao (2011).

\begin{lem} \label{lem:1}
Denote $\varepsilon_{i}(\tau)=y_{i}-\xi_{i}(\tau)$, and let $\vartheta^*=(\rho,\beta,\Theta,\zeta)$ be a parameter vector in $\mathcal{R}\times\mathcal{B}\times\mathcal{S}\times\mathcal{Z}$. Let
\begin{equation}
\delta=\begin{pmatrix}
\delta_{\rho} \\
\delta_{\beta} \\
\delta_{\Theta} \\
\delta_{\zeta} \\
\end{pmatrix}
=\begin{pmatrix}
\sqrt{n}(\hat\rho(\tau)-\rho(\tau)) \\
\sqrt{n}(\hat\beta(\tau)-\beta(\tau)) \\
\sqrt{n}(\hat\Theta(\tau)-\Theta(\tau)) \\
\sqrt{n}(\hat\zeta(\tau)-\zeta(\tau)) \\
\end{pmatrix}.
\end{equation}
Under Assumptions 1-3, we have
\begin{align*}
&\underset{\vartheta^*}
{\sup}\frac{1}{n}\bigg|\sum_{i=1}^{n}\bigg[\rho_{\tau}\bigg(\varepsilon_{i}(\tau)
-\frac{d_{i}\delta_{\rho}}{\sqrt{n}}-\frac{X_{i}\delta_{\beta}}{\sqrt{n}}
-\frac{\Pi_{i}\delta_{\Theta}}{\sqrt{n}}-\frac{\omega_{i}\delta_{\zeta}}{\sqrt{n}}\bigg)
-\rho_{\tau}(\varepsilon_{i}(\tau))\\
&-E\bigg[\rho_{\tau}\bigg(\varepsilon_{i}(\tau)-\frac{d_{i}\delta_{\rho}}{\sqrt{n}}
-\frac{X_{i}\delta_{\beta}}{\sqrt{n}}-\frac{\Pi_{i}\delta_{\Theta}}{\sqrt{n}}
-\frac{\omega_{i}\delta_{\zeta}}{\sqrt{n}}\bigg)
-\rho_{\tau}(\varepsilon_{i}(\tau))\bigg]\bigg]\bigg|=o_p(1).
\end{align*}
\end{lem}

\subsection*{1\quad Proof of Theorem \ref{th:1}}

\textbf{Proof.} Firstly, following Chernozhukov and Hansen (2006), $\vartheta_1^*(\tau)=(\rho(\tau), \beta(\tau), \Theta(\tau))$ uniquely solves the problem for each $\tau$.

To prove the consistency of the parameter, we need to show that under Assumptions 1-3, $\hat\vartheta^*_1(\tau)=\vartheta^*_1(\tau)+o_p(1)$. Let
$$
\mathcal{P}: \vartheta^*_1\mapsto\rho_\tau(y-\rho D-X\beta-\Pi\Theta),
$$
and $\mathcal{P}$ is continuous. Under condition Lemma \ref{lem:1}, we have that $\|\hat\vartheta^*(\rho,\tau)-\vartheta^*(\rho,\tau)\|\stackrel{P}{\rightarrow}0$ for $\vartheta^*=(\rho,\beta,\Theta,\zeta)$, which implies that $\|\|\hat\zeta(\rho,\tau)\|-\|\zeta(\rho,\tau)\|\|\stackrel{P}{\rightarrow}0$. By Corollary 3.2.3 in van der Vaart and Wellner (1996), we have $\|\hat\rho(\tau)-\rho(\tau)\|\stackrel{P}{\rightarrow}0$. Therefore, $\|\hat\beta(\hat\rho(\tau),\tau)-\beta(\tau)\|\stackrel{P}{\rightarrow}0$, $\|\hat\Theta(\hat\rho(\tau),\tau)-\Theta(\tau)\|\stackrel{P}{\rightarrow}0$,
and $\|\hat\zeta(\hat\rho(\tau),\tau)-0\|\stackrel{P}{\rightarrow}0$. Hence, $\|\hat\vartheta^*(\tau)-\vartheta^*(\tau)\|\stackrel{P}{\rightarrow}0$.

Using Minkowski inequality, Assumptions 2(ii) and 3(iv), we know $\sup_{u\in\mathcal{U}}\|\gamma(u,\tau)-\Pi\Theta(\tau)\|=O((k_n+h+1)^{-r})$, hence
\begin{align*}
&\quad\sup_{u\in\mathcal{U}}\|\hat\gamma(u,\tau)-\gamma(u,\tau)\|\\
&\leq\sup_{u\in\mathcal{U}}\|\Pi(\hat\Theta(\tau)-\Theta(\tau))\|
+\sup_{u\in\mathcal{U}}\|\Pi\Theta(\tau)-\gamma(u,\tau)\| \\
&=O(\sqrt{k_n+h+1})\cdot o_p(1)+O_p((k_n+h+1)^{-r})\\
&=O_p((k_n+h+1)^{-r}).
\end{align*}
\begin{flushright}
$\square$
\end{flushright}

\subsection*{2\quad Proof of Theorem \ref{th:2}}

For any $\hat\rho(\tau)\stackrel{P}{\rightarrow}\rho(\tau)(\delta_{\rho}\stackrel{P}{\rightarrow}0)$, we
can write the objective function defined in equation \eqref{ob:iv} as
$$
R_{IV}=\sum_{i=1}^{n}\bigg[\rho_{\tau}\bigg(\varepsilon_{i}(\tau)-\frac{d_{i}\delta_\rho}{\sqrt{n}}
-\frac{X_{i}\delta_{\beta}}{\sqrt{n}}-\frac{\Pi_{i}\delta_{\Theta}}{\sqrt{n}}
-\frac{\omega_{i}\delta_{\zeta}}{\sqrt{n}}\bigg)-\rho_{\tau}(\varepsilon_{i}(\tau)\bigg]
$$
where $\varepsilon_{i}(\tau)=y_{i}-\xi_{i}(\tau)$, $\xi_{i}(\tau)=\rho(\tau)d_{i}+X_{i}\beta(\tau)+\Pi_{i}\Theta(\tau)+\omega_{i}\zeta(\tau)$, and
$$
\delta=\begin{pmatrix}
\delta_{\rho} \\
\delta_{\beta} \\
\delta_{\Theta} \\
\delta_{\zeta}
\end{pmatrix}
=\begin{pmatrix}
\sqrt{n}(\hat\rho(\tau)-\rho(\tau)) \\
\sqrt{n}(\hat\beta(\tau)-\beta(\tau)) \\
\sqrt{n}(\hat\Theta-\Theta) \\
\sqrt{n}(\hat\zeta(\tau)-\zeta(\tau))
\end{pmatrix}.
$$
Let $\varphi_\tau(u)=\tau-I(u<0)$ and
$$
G(\delta_{\rho},\delta_{\beta},\delta_{\Theta},\delta_\zeta)=\frac{-1}{\sqrt{n}}\sum_{t=1}^{\top}
\varphi_\tau\bigg(\varepsilon_{i}(\tau)-\frac{d_{i}\delta_\rho}{\sqrt{n}}-\frac{X_{i}\delta_{\beta}}{\sqrt{n}}
-\frac{\Pi_{i}\delta_{\Theta}}{\sqrt{n}}-\frac{\omega_{i}\delta_{\zeta}}{\sqrt{n}}\bigg).
$$
Let
$$
\sup\|G(\delta_{\rho},\delta_{\beta},\delta_{\Theta},\delta_\zeta)-G(0,0,0,0)
-\mathbb{E}[G(\delta_{\rho},\delta_{\beta},\delta_{\Theta},\delta_\zeta)-G(0,0,0,0)]\|=o_p(1).
$$
Expanding $G$, we obtain
\begin{align*}
&\quad\mathbb{E}[G(\delta_{\rho},\delta_{\beta},\delta_{\Theta},\delta_\zeta)-G(0,0,0,0)] \\
&=\frac{1}{\sqrt{n}}\sum_{i=1}^{n}
\tilde{X}^{\top}_{i}f_{i}(\xi_{i}(\tau))\bigg[\frac{d_{i}\delta_{\rho}}{\sqrt{n}}+\frac{X_{i}\delta_{\beta}}{\sqrt{n}}
+\frac{\Pi_{i}\hat\delta_{\Theta}}{\sqrt{n}}
+\frac{\omega_{i}\delta_{\zeta}}{\sqrt{n}}\bigg]+o_p(1).
\end{align*}
where $\tilde{X}=[X,\Pi,E]$. Obviously, $G(\hat\delta_{\rho},\hat\delta_{\beta},\hat\delta_{\Theta},\hat\delta_\zeta)\rightarrow0$, $\mathbb{E}[G(\delta_{\rho},\delta_{\beta},\delta_{\Theta},\delta_\zeta)-G(0,0,0,0)]=-G(0,0,0,0)$, i.e., the last equation has the following equivalent expression:
$$
\frac{1}{\sqrt{n}}\sum_{i=1}^{n}\tilde{X}^{\top}_{i}\varphi_\tau(\varepsilon_{i}(\tau))=
\frac{1}{\sqrt{n}}\sum_{i=1}^{n}
\tilde{X}^{\top}_{i}f_{i}(\xi_{i}(\tau))\bigg[\frac{d_{i}\delta_{\rho}}{\sqrt{n}}+\frac{X_{i}\delta_{\beta}}{\sqrt{n}}
+\frac{\Pi_{i}\hat\delta_{\Theta}}{\sqrt{n}}
+\frac{\omega_{i}\delta_{\zeta}}{\sqrt{n}}\bigg]+o_p(1).
$$
Letting $\delta_{\eta}=(\delta^\top_{\beta},\delta^\top_{\Theta},\delta_{\zeta})^\top$, we write the equation above as:
$$
\frac{1}{\sqrt{n}}\sum_{i=1}^{n}
\tilde{X}^{\top}_{i}f_{i}(\xi_{i}(\tau))\bigg[\frac{d_{i}\delta_{\rho}}{\sqrt{n}}+\frac{\tilde{X}_{i}
\delta_{\eta}}{\sqrt{n}}\bigg]
=\frac{1}{\sqrt{n}}\sum_{i=1}^{n}\tilde{X}^{\top}_{i}\varphi_\tau(\varepsilon_{i}(\tau)).
$$

Alternatively, using more convenient notation, we write the last expression as:
$$
\mathbf{J}_{\rho}\delta_{\rho}+\mathbf{J}_{\eta}\delta_{\eta}=\mathbb{J}_{\phi},
$$
where $\mathbf{J}_{\eta}=\underset{n\rightarrow\infty}{\lim}\tilde{X}^{\top}\Omega\tilde{X}$, $\mathbf{J}_{\rho}=\underset{n\rightarrow\infty}{\lim}\tilde{X}^{\top}\Omega D$, $\mathbb{J}_{\phi}$ is a mean zero r.v. with
covariance $\tau(1-\tau)\tilde{X}^{\top}\tilde{X}$, $\Omega=diag(f_{i}(\xi_{i}(\tau)))$ and $\Phi_{\tau}$ is a $NT$-vector $(\phi_\tau(\varepsilon_{i}(\tau)))$.

Letting $[\bar{\mathbf{J}}_{\beta},\bar{\mathbf{J}}_{\Theta},\bar{\mathbf{J}}_\eta]$ be a
conformable partition of $\mathbf{J}^{-1}_{\eta}$ as in Galvao (2011) and Chernozhukov and Hansen (2006) yields
$\hat\delta_{\beta}=\bar{\mathbf{J}}^{\top}_{\beta}(\mathbb{J}_\phi-\mathbf{J}_{\rho}\delta_{\rho})$,
$\hat\delta_{\Theta}=\bar{\mathbf{J}}^{\top}_{\Theta}(\mathbb{J}_\phi-\mathbf{J}_{\rho}\delta_{\rho})$,
and $\hat\delta_{\eta}=\bar{\mathbf{J}}^{\top}_{\eta}(\mathbb{J}_\phi-\mathbf{J}_{\rho}\delta_{\rho})$.
Letting $H=\bar{\mathbf{J}}^{\top}_\eta A\bar{\mathbf{J}}_\eta$ as in Chernozhukov and Hansen (2006) gives
$\hat\delta_{\rho}=K\mathbb{J}_\phi$, where $K=(\mathbf{J}_{\rho}^{\top}H\mathbf{J}_{\rho})^{-1}\mathbf{J}_{\rho}^{\top}H$.
Replacing it in the previous expression, $\hat\delta_{\eta}=\bar{\mathbf{J}}^{\top}_{\eta}(\mathbb{J}_\phi-\mathbf{J}_{\rho}\delta_{\rho})
=\bar{\mathbf{J}}^{\top}_{\eta}(I-\mathbf{J}_{\rho}(\mathbf{J}_{\rho}^{\top}H\mathbf{J}_{\rho})^{-1}
\mathbf{J}_{\rho}^{\top}H)\mathbb{J}_\phi=\bar{\mathbf{J}}^{\top}_{\eta}M
\mathbb{J}_\phi$, where $M=I-\mathbf{J}_{\rho}(\mathbf{J}_{\rho}^{\top}H\mathbf{J}_{\rho})^{-1}
\mathbf{J}_{\rho}^{\top}H$. Due to the  invertibility of $\mathbf{J}_{\rho}\bar{\mathbf{J}}_{\eta}$,
$\hat\delta_\eta=\mathbf{0}\times O_p(1)+o_p(1)$. Similarly, substituting back $\delta_{\rho}$, we obtain that $\hat\delta_{\beta}=\bar{\mathbf{J}}^{\top}_{\beta}M\mathbb{J}_\phi$ and $\hat\delta_{\Theta}=\bar{\mathbf{J}}^{\top}_{\Theta}M\mathbb{J}_\phi$.
By the regularity conditions, we have that
$$
\begin{pmatrix}
\hat\delta_{\rho}(\rho_n,\tau)\\
\hat\delta_{\beta}(\rho_n,\tau) \\
\hat\delta_{\Theta}(\rho_n,\tau)
\end{pmatrix}=
\begin{pmatrix}
\sqrt{n}(\hat{\rho}(\rho_n,\tau)-\rho(\tau))\\
\sqrt{n}(\hat{\beta}(\rho_n,\tau)-\beta(\tau)) \\
\sqrt{n}(\hat{\Theta}(\rho_n,\tau)-\Theta(\tau))
\end{pmatrix}\rightsquigarrow\mathcal{N}(\mathbf{0},J^{\top}SJ).
$$
where $J=(K^{\top},L^{\top}_1,L^{\top}_2)$, $L_1=\bar{J}^\top_{\beta}M$, $L_2=\bar{J}^\top_{\Theta}M$.

Divide $L_2$ as $[L_2^{(1)\top},\cdots,L_2^{(q)\top}]^\top$. Let $\Pi^{(l)}=[\Pi^{(l)\top}_1,\cdots,\Pi^{(l)\top}_n]^\top$, where $\Pi^{(l)}_i=Z_{il}\pi_{k_{n}}^{\top}(U_i)$. Then we have $\hat\delta_{\gamma l}=\hat\gamma_l(u,\tau)-\gamma_l(u,\tau)=\Pi^{(l)}\hat\delta_{\theta l}+\Pi^{(l)}\theta_l(\tau)-\gamma_l(u,\tau)
\approx\Pi^{(l)}\hat\delta_{\theta l}=\Pi^{(l)}L_2^{(l)}\mathbb{J}_\phi$. By the regularity conditions, we have that
$$
\sqrt{n}(\hat{\gamma}_l(\rho_n,u,\tau)-\gamma(u,\tau))\rightsquigarrow\mathcal{N}(\mathbf{0},L_3^{(l)}SL_3^{(l)\top}).
$$
where $L_3^{(l)}=\Pi^{(l)}L_2^{(l)}$.
\begin{flushright}
$\square$
\end{flushright}

Theorem \ref{th:5} can be easily proved, and proofs of Theorem \ref{th:3} and \ref{th:4} are similar as those of Theorem 3 and 4 in Wang et al. (2009). Thus proofs of Theorem \ref{th:5}, \ref{th:3} and \ref{th:4} are omitted here.

\end{document}